# Landau Modeling of Dynamical Nucleation of Martensite at Grain Boundaries under Local Stress

Guanglong Xu[1], Chuanyun Wang[1], Juan Ignacio Beltrán[1], Javier LLorca[1,2], Yuwen Cui[1†]

[1]IMDEA Materials Institute, C/Eric Kandel 2, 28906 – Getafe, Madrid, Spain.

[2]Department of Materials Science, Polytechnic University of Madrid, E. T. S. de Ingenieros de Caminos, 28040 – Madrid, Spain

*Abstract*

The dynamical nucleation of martensite in polycrystals is simulated by means of Lagrange-Rayleigh dynamics with Landau energetics, which is capable of obtaining the local stress as a result of the interplay of the potential of transformation and external loadings. By monitoring the spatio-temporal distribution of the strain in response to the local stress, we demonstrate that the postcursors, high angle grain boundaries and triple junctions act as favorable heterogeneous nucleation sites corresponding to different loading and cooling conditions, and predict the phase diagram of the nucleation mode of martensite.

## 1. Introduction

A preference for heterogeneous nucleation in Martensitic Transformations (MT) has been evidenced by a wide body of experimental explorations on steels [1-3], shape-memory alloys [4-6], ceramics [7-8], etc. The heterogeneous processes of nucleation and subsequent growth at various defects present distinct microstructure [9-10], diverse transformation pathways [11-12], and hence give rise to different properties of materials [13-14]. To understand the mechanisms underlying the observed phenomena, different models of martensitic nucleation have been developed for different defects like dislocations [15-21] and grain boundaries [22-31] in perspectives of thermodynamics, kinetics and crystal lattice dynamics. These models are at least classified into two categories considering the classical and non-classical nucleation paths [32-33]. The models describing classical nucleation paths believe that the embryos of martensites are formed via the dislocation dissociation mechanism and thereafter increase in size by motions of interfaces. During these processes, the nuclei have a fixed structure and/or the same composition as those of fully formed martensite variants. This is the case such that the driving force and kinetics of martensitic nucleation can be discerned by evaluating the time evolution of individual energy contributions from misfit elastic energy, chemical energy in bulk (composition-dependent lattice stability) and surface energy. Pioneered by Cohen [15], Kaufman [16], Christian [17], and consolidated by Olson [18, 32-35] *et al*., the model for classical paths has been developed into a self-consistent and unified interpretation to heterogeneous martensitic nucleation on a basis of dislocation theories, thermodynamics and crystallography of martensitic transformation, and it has been widely applied to the martensite nucleation in steels and a few shape memory alloys, where the dislocation plays the primary role.

In contrast, the models for non-classical martensitic nucleation paths describe a continuous change in structure and/or composition in a finite temperature region. Among the most prominent is the dynamical nucleation model [36], where the Landau-type free energy, incorporating physical nonlinearity describing the symmetry breaking in MT and nonlocal terms accounting for the long-range interactions of elastic oscillators, is formulated in a frame of Lagrangian-Rayleigh (LR) or Bales-Gooding (BG) [37-38] dynamics to present a physics scenario of 'elastic solitary wave' for martensitic nucleation [37-45]. Different from the classical models, the dynamical nucleation method is able to investigate the complete process of martensitic interface motion and interpret the phenomena at early stage of MT such as the formation of tweeds, dynamical twinning, autocatalysis, etc. Thus, it is helpful and has persistently been used in studying the homogeneous coherent nucleation of improper and weak proper MTs in shape memory alloys, which arise from lattice softening and/or weak distortion near the critical point. Meanwhile, further uses of the model towards studying heterogeneous nucleation in weak proper MT have been performed. The fundamental physical picture of heterogeneous dynamical nucleation model is depicted as follows: The interaction of the defects with the material allows a spatiotemporally varying potential of phase transformation and gives rise to local stress field in the correct tensor expression and magnitude to exert influence on MT [46]. Clapp *et al*. [39] incorporated for the first time the Ginzburg-Landau (GL) phenomenological theory to determine the 'spinodal' strains during martensite nucleation by introducing the interaction of a single 'misfit' planar defect with the host. They denominated this type of nucleation as 'localized soft mode'. Later, the initial concept of heterogeneous dynamical nucleation was further consolidated with the efforts of Cao [40-41], Reid [36, 43-45], Chu [47], van Zyl [48-49] and Gröger [50] *et al*. Based on the dynamical twinning [37], Reid *et al*. [43] extensively analyzed the dynamical nucleation with a predefined local strain field (as an analog of heterogeneous nucleus) along with generalized boundary conditions. Their results demonstrated that the variation of morphology in MT, e.g. the presence of twin or single-domain martensite, depends on the quenching temperature and dissipation, while it is independent of the boundary conditions. Afterwards, van Zyl [48-49] *et al*. inferred that the surface nucleation, a mode by which the intrinsic inhomogeneous strain field coupled with the boundaries leads to a lower saddle point on the transformation energy surface, would always be preferred over homogeneous bulk nucleation. Cao [40-41] introduced a planar defect with a predefined stress field (instead of aforementioned predefined strain field) and ascertained the critical stresses and the effective temperature for the onset of MT in the square-to-rectangular ferroelastic materials. A similar formalism of heterogeneous nucleation was also presented by Gooding *et al*. [42], yet the latter focused on the concentration of the local strain at the defect arising from undercooling. It was further applied by Reid *et al.* [45] to the Eshelby's inclusion problem where the functional forms or the values of the local stress field due to the presence of the inclusion become available.

Other than the aforementioned works which focus on the MT induced by the single planar defect with a predefined stress or strain, few works of heterogeneous dynamical nucleation have been applied to grain boundary with more complicated geometry of defects. It is in comparison with the great achievements of modern phase-field methods on this topic [22-31]. The phase-field models have advantages to represent the perfect martensitic microstructure by taking a mutually advantageous conjunction of interface motion and clarified energy contributions in models for classical paths and the continuously varying order parameters in models for non-classical paths, which as a whole describe a hybrid transformation path. However, the over-damped dynamics in phase field method encounters difficulties in accounting for the correct dynamic behaviors for the formation of precursors [51] and the near sound velocity of interface propagation in weak proper MT [38]. This is due to the fact that these behaviors are the consequence of competition among inertia, damping and undercooling, as well as the minimization of kinetic energy [38]. The interplaying of these effects in a strained system undergoing weak first-order MT reflects the under-damped dynamics of elastic oscillators [36]. Therefore, the development of the complete dynamical nucleation model in polycrystals remains open. Recently, Ahluwalia *et al*. [52] extended strain based Landau energetics and LR dynamics to investigate the grain size effects on the MT in nano-polycrystalline shape memory alloys, especially when the MT is inhibited around grain boundaries in nano-polycrystals. However, the phenomenon where the alignment of martensite variants rotate at certain angles with respect to the grain orientations in individual grains, generally observed by microscopy, is unnoticeable in their simulated morphology. This is because they adopted global stresses in the dissipative force balance equation which led to the mechanical equilibrium of global stresses rather than local transformation stresses. Therefore, we modified the existing polycrystalline MT model by replacing the global stresses with local transformation stresses and then rewrote them in global variables, i.e. displacement gradients in global coordinates [53]. The updated model is not only consistent with kinematic compatibility in Phenomenological Theory of Martensite Crystallography (PTMC), but also has the ability to explicitly explore the spatiotemporal distribution of local stress during the MT. It also allows us to generalize the conceptual dynamical nucleation theory within our updated model to simulate the heterogeneous martensitic nucleation in polycrystals. The present simulations aim at extending the application of dynamical nucleation model and at gaining an in-depth understanding of the collective (competitive and/or cooperative) effects of the martensitic nucleation and subsequent growth assisted by grain boundaries and external loadings following the non-classical transformation paths. In turn, it allows evaluating the ability of different defects/positions as martensitic nucleation sites originating from the interaction of intrinsic polycrystalline defects with the applied loading in a weak proper MT.

## 2. Model and Numerical Implementation

The dynamical nucleation of MT induced by grain boundary in polycrystals can be achieved once the predefined planar defects in Refs. [40-42] are specified as the grain boundaries and triple junctions. The grain boundaries and triple junctions can perceive the applied stress and feel back to couple with Landau transformation potential, which leads to a local stress field and modifies the transformation temperature. Similar to Cao's work on single crystal [40], we assume that 1) the inhomogeneous distribution of stress is produced by defects, specifically in this work, the local stress fields around the grain boundaries and triple junctions, rather than the defects themselves; 2) the interface between austenite and martensite is coherent which is imposed by the compatibility condition; 3) the displacement at grain boundaries is continuous as required by the displacement-based LR dynamics [38]. It is inferred that martensitic nucleation against free surfaces and incoherent grain boundaries is not taken into consideration. It is also noted that we are not intending to obtain various morphologies by tuning the damping parameters as what conventional dynamical nucleation modeling did, rather, we focus on how intrinsic MT potential involving grain boundaries and applied stresses leads to different nucleation modes or sites.

There are two critical issues to be addressed in our Landau modeling of the dynamical nucleation in polycrystals: the Landau free energy in the expansion of order parameters and the deterministic dynamic equations. In the single crystal model, non-linear Landau free energy is developed based on the linearized strain tensor with the component $e_{ij} = \frac{1}{2}\left(u_{i,j} + u_{j,i}\right)$, where $u_i$ denotes the local displacement, and $u_{i,j} = \partial u_i / \partial x_j$ is the displacement gradient in the intragranular coordinates. The symmetry-adapted linear strains $e_k$, written as the linear combination of the strain components, are selected as the Order Parameters (OP). Now we consider a MT in polycrystals. The degrees of freedom in a system undergoing MT are the global displacements, $U_i$, and they should be connected with the OP strains by introducing an additional variable $\theta(\vec{r},t)$ describing the grain orientation field. As such, the local strains and symmetry-adapted OP strains are the functional of the global displacement gradients, viz. $e_{ij}(U_{I,J})$ and $e_k(U_{I,J})$. For the sake of clarity, the variables in global coordinates are denoted by capital letters while those in local coordinates by lowercase. The specific expressions of free energy and dynamic equations for the simulation of 2D square-to-rectangular martensitic nucleation are detailed as follows.

## 2.1. Landau free energy

The free energy functional of the system is written as the spatial integral of free energy density $f$,

$$F = \int f d\vec{r} = \int d\vec{r} \left( f_{elastic} + f_{grain} + f_{load} \right), \quad \ldots\ldots(1)$$

while $f$ is intuitively defined as the summation of three energy density contributions: the transformation energy density in elasticity $f_{elastic}$, the energy of a polycrystalline microstructure $f_{grain}$, and the energy due to the applied load $f_{load}$.

In our model, the grain orientation field of polycrystals is described by the conventional phase field method. It defines the spatially distributed grain orientation $\theta\left[\vec{\eta}_i(\mathbf{r},t)\right]$ with respect to a set of arbitrarily selected intragranular coordinates which should serve as the referential global coordinates. The grain orientation field is expressed as

$$\theta(\vec{\eta},\vec{\mathbf{r}}) = \frac{\theta_{max}}{Q-1}\left[\frac{\sum_{i=1}^{Q}(i-1)\eta_i(\vec{\mathbf{r}},t)}{\sum_{i=1}^{Q}\eta_i(\vec{\mathbf{r}},t)}\right] \quad \ldots\ldots(2)$$

where, $\vec{\eta}(\eta_1, \eta_2 \ldots \eta_Q)$ is a vectorial indicator, by which a given grain orientation corresponds to one component $\eta_i$ being positive nonzero while the remaining components are equal to zero [52]. The energy contribution $f_{grain}$ quantifies the spatial inhomogeneity in crystal orientation and can be written as

$$f_{grain} = \varpi\left\{\sum_{i=1}^{Q}\left[\frac{a_2}{2}\eta_i^2 + \frac{a_3}{3}\eta_i^3 + \frac{a_4}{4}\eta_i^4\right] + \frac{a_c}{2}\sum_{i=1}^{Q}\sum_{j\neq i}^{Q}\eta_i^2\eta_j^2 + \sum_{i=1}^{Q}\frac{\kappa_{grain}}{2}(\nabla\eta_i)^2\right\} \quad \ldots\ldots(3)$$

where, the coefficients are $a_2$, $a_3 < 0$ and $a_4$, $a_c > 0$ to ensure a potential with $Q$ degenerate minima, $(\eta_1, 0, \ldots, 0)$, $(0, \eta_2, \ldots, 0)$, etc. up to $(0, 0, \ldots, \eta_Q)$ $(\eta_i > 0)$, $\varpi$ is an adjustable parameter to mediate the magnitude of grain boundary energy density. It's worth noting that the polycrystalline structure in our model is simply described by the spatial distribution of different grain orientations, thus the internal stress at grain boundaries are absent in comparison with the real polycrystals.

The elastic transformation energy density $f_{elastic}$ can be further divided into three parts, i.e., local transformation energy density $f_{local}$, energetic contribution originating from non-OP strains $f_{non-OPs}$, and gradients of OP strains $f_{grad}$. The first term $f_{local}$ describes the two degenerated martensite variants due to the symmetry breaking in square-to-rectangular MT, and is expressed in the polynomial expansion of the symmetry-adapted deviatoric strain, $e_2$, up to sixth order,

$$f_{local} = \frac{A_{20}}{2}\left(\frac{T-T_c}{T_m-T_c}\right)e_2^2 + \frac{A_4}{4}e_2^4 + \frac{A_6}{4}e_2^6 , \quad \ldots\ldots(4)$$

where, $A_2 = A_{20}\left(\frac{T-T_c}{T_m-T_c}\right)$ is temperature dependent deviatoric modulus expressed in linear combination of the elastic constants as $A_2 = c_{11}(T) - c_{12}(T)$. $A_4$ and $A_6$ are related to higher order nonlinear elastic constants. $T_c$ and $T_m$ are critical point and MT starting temperature, respectively. The OP strain $e_2 = (e_{xx} - e_{yy})/\sqrt{2}$ has to be rewritten as the derivation of the global displacements in global coordinates, i.e.

$$e_2 = [u_{x,x}(U,\theta) - u_{y,y}(U,\theta)]/\sqrt{2}, \quad \ldots\ldots(5.1)$$

$$= [\cos 2\theta (U_{X,X} - U_{Y,Y}) + \sin 2\theta (U_{X,Y} + U_{Y,X})]/\sqrt{2}, \quad \ldots\ldots(5.2)$$

where, $\theta$ is the angle of grain orientation, see Eq.(2). There are two non-OPs, i.e. dilatational strain $e_1 = (e_{xx} + e_{yy})/\sqrt{2} = (U_{X,X} + U_{Y,Y})/\sqrt{2}$ and the shear strain $e_3 = (e_{xy} + e_{yx})/2 = -\dfrac{\sin(2\theta)}{2}(U_{X,X} - U_{Y,Y}) + \dfrac{\cos(2\theta)}{2}(U_{X,Y} + U_{Y,X})$ to construct the $f_{non\text{-}OPs}$ in square-to-rectangular MT as:

$$f_{non-OPs} = \frac{A_1}{2} e_1^2 + \frac{A_3}{2} e_3^2, \quad \ldots\ldots(6)$$

where, $A_1 = c_{11} + c_{12}$ is the bulk modulus in 2D, $A_3 = 4c_{44}$ is the shear modulus in the isotropic approximation. The gradient energy density $f_{grad}$ is the square of the OP strain gradient, i.e.

$$f_{grad} = g(\nabla e_2)^2, \quad \ldots\ldots(7)$$

and $g$ is the gradient coefficient.

Since we are interested in a simple uniaxial stress in this work, the free energy contribution due to the applied stress $\sigma_{app}$ yields

$$f_{app} = -\sigma_{app} e_{xx}, \quad \ldots\ldots(8)$$

Thus, the free energy is finally expressed in global coordinates as

$$F = \int d\vec{r} \begin{bmatrix} \dfrac{A_1}{2} e_1^2(\vec{r},\theta) + \dfrac{A_{20}}{2}\left(\dfrac{T-T_c}{T_m-T_c}\right) e_2^2(\vec{r},\theta) + \dfrac{A_4}{4} e_2^4(\vec{r},\theta) + \dfrac{A_6}{6} e_2^6(\vec{r},\theta) + \dfrac{A_3}{2} e_3^2(\vec{r},\theta) \\ + g(\nabla e_2(\vec{r},\theta))^2 + f_{grains}(\theta) - \sigma_{app} e_{xx}(\vec{r},\theta) \end{bmatrix} \ldots\ldots(9)$$

## 2.2. Dynamic equations

The evolution of grain orientation and displacement fields are treated differently in this work. The time dependent Ginzburg-Landau equation

$$\frac{\partial \vec{\eta}(\vec{r},t)}{\partial t} = -L_\eta \frac{\delta f_{grain}}{\delta \vec{\eta}(\vec{r},t)} \ldots\ldots(10)$$

is used to describe the evolution of the order parameter field $\vec{\eta}(\vec{r},t)$ of the grain orientations with $L_\eta$ as kinetic coefficient for interface mobility.

When the spatiotemporal evolution of displacement and strain fields is considered, a variation method of Lagrange mechanics is utilized. We incorporate the kinetic energy density $T = \rho \dot{\mathbf{U}}^2/2$ and the free energy density $f$ to construct the Lagrangian $\mathscr{L} = \int_V d\mathbf{r}\{T - f\}$. By taking the variation of the Lagrangian with respect to the displacement $\mathbf{U}$ and velocity $\dot{\mathbf{U}}$ and explicitly substituting the Lagrangian into the Lagrange-Rayleigh equations $\frac{d}{dt}\frac{\delta \mathscr{L}}{\delta \dot{U}_i} - \frac{\delta \mathscr{L}}{\delta U_i} = -\frac{\delta \mathscr{R}}{\delta \dot{U}_i}$, the Lagrange-Rayleigh equations yield the general equations of motion

$$\rho \ddot{U}_i = \frac{\partial}{\partial X_j}\left\{\frac{\partial f}{\partial e_k}\frac{\partial e_k}{\partial e_{pq}}\frac{\partial e_{pq}}{\partial U_{i,j}} - \nabla \cdot \frac{\partial f}{\partial \nabla e_k}\frac{\partial \nabla e_k}{\partial e_{pq}}\frac{\partial e_{pq}}{\partial U_{i,j}} + \frac{\partial R}{\partial \dot{e}_k}\frac{\partial \dot{e}_k}{\partial \dot{e}_{pq}}\frac{\partial \dot{e}_{pq}}{\partial \dot{U}_{i,j}}\right\}, \quad \ldots\ldots(11)$$

where, $\rho$ is density; $\mathscr{R}$ and R are Rayleigh dissipation and dissipation density, written as the functional of strain rates to describe the friction in the system. If the isotropic approximation is adopted for the Rayleigh dissipation density with viscosity $\gamma$, and the local stress expressed in global coordinates is defined as

$$\sigma_{ij} = \frac{\delta F}{\delta U_{i,j}} = \frac{\partial f}{\partial e_k}\frac{\partial e_k}{\partial e_{pq}}\frac{\partial e_{pq}}{\partial U_{i,j}} - \nabla \cdot \frac{\partial f}{\partial \nabla e_k}\frac{\partial \nabla e_k}{\partial e_{pq}}\frac{\partial e_{pq}}{\partial U_{i,j}}, \quad \ldots\ldots(12)$$

the dynamics of displacement fields is given by dissipative force balance equations

$$\rho \ddot{U}_i = \frac{\partial \sigma_{ij}}{\partial X_j} + \gamma \nabla^2 \dot{U}_i, \quad \ldots\ldots(13)$$

Differing from Ref [52], the local stresses in our model enter dissipative force balanced equations via the chain rule of functional differentiation of free energy with respect to global displacement gradients. By numerical solving Eqs. (10), (12) and (13), the spatiotemporal evolution of displacements, strains and grain orientations are obtained. Finally, the viscous term drives the system to the mechanical equilibrium state.

## 2.3. Numerical Implementation

Our simulations were carried out using a Fast Fourier Transform (FFT)-based spectral method under periodic boundary conditions in a 256×256 grid with 5 grain orientations of 0°, 7.5°, 15°, 22.5° and 30° (=$\theta_{max}$), respectively (Fig. 1-a). For the formation of the polycrystalline structure, we used the parameters, $a_2$=-1, $a_3$=-1, $a_4$=1, and $a_c$=2. $k_{grain}$ was adjustable to produce appropriate configuration. Since we took a Fe-Pd ferroelastic alloy as an example in our simulation, the following materials parameters were chosen based on the experimental results of elastic constants: $A_1$ = 333.91 GPa, $A_{20}$ = 11.5 GPa, $A_3$ = 282 GPa, $A_4$ = -1.7*10$^3$ GPa, $T_m$ = 295 K and $T_c$ =270 K [50,54-55]. Note that the temperature dependence of the elastic constant $A_2$ in experiments and linear fitting was well represented by the parameter $A_2 = A_{20}\frac{T-T_c}{T_m-T_c}$, while the other elastic parameters are constant with temperature for simplicity. $A_6$ was determined from $A_{20}$, $A_4$, and quenching temperature $T$, satisfying the constraints of Landau polynomial for first-order transformation [56]. The value of the gradient coefficient was set to $g$=3.0*10$^{-8}$ N [52,57] based on microstructural observations, while the dissipation coefficient was estimated as 0.015N s/m$^2$ [57]. Based on these parameters, the interface energy density of full twinned martensites can be estimated around 0.5J/m$^2$ from multi-wells degenerated Landau free energy near the critical point*.

*If the viscous term is removed, the conservation equation of motion utilized in our simulation describes a physical picture in which the decrease of local free energy in Landau potential triggers the formation of the habit planes and/or twin boundaries during the martensitic nucleation and drives the interface propagation during martensite growth. In fact, the existence of a viscous term minimizes the finite kinetic energy density.

The interface energy density in our simulation (in Subsection 3.1) has the correct order of magnitude, consistent with theoretical estimations and first principles calculations [58-59]. To tackle the numerical instability, both the materials and the simulation parameters were subjected to a normalization [60], wherein the spatial and energetic variables were rescaled by introducing the normalization factors $d_0$=7*10$^{-9}$m and $f_0$ ~10$^{10}$J, respectively; such that $\tilde{x}=x/d_0$, and $\tilde{A}_i = A_i/f_0$. The time variable was rescaled as $\tilde{t} = t/t_0 = t/\sqrt{\rho d_0^2/f_0}$ and the rescaled viscosity is $\tilde{\gamma} = \gamma/\left(d_0\sqrt{f_0\rho}\right)$. At the beginning of the simulation, the random fluctuations of displacements $\zeta_{u_i}$ obtained from a Gaussian noise were introduced. The mean value and correlation of fluctuations

satisfy

$$\langle \zeta_{u_i}(\mathbf{r},t) \rangle = 0, \quad \ldots\ldots(14)$$

$$\langle \zeta_{u_i}(\mathbf{r},t) \zeta_{u_j}(\mathbf{r}',t') \rangle = 2\gamma k_B T \rho \delta(\mathbf{r}-\mathbf{r}')\delta(t-t') \ldots\ldots(15)$$

More details of the model and of the numerical implementation can be found in Refs. [53,60-61].

## 3. Results and Discussion

The martensitic nucleation and the relevant microstructure evolution are obtained from the simulations under the conditions of (1) different applied stresses at a fixed temperature, (two cases of martensitic nucleation under applied stresses of 50 MPa and 500 MPa are selected as the examples to be discussed in detail), (2) different temperatures with a constant applied stress, and (3) the combination of varying stresses and temperatures. These external processing conditions interacting with intrinsic MT energy lead to different scenarios of martensitic nucleation.

### *3.1 Heterogeneous nucleation under an applied stress of 50 MPa*

To examine the MT in the vicinity of grain boundaries, the system was first quenched to $T = 275K$ ($T_c<T<T_m$), followed by an isothermal simulation under a constant applied stress of $\sigma_{app}$= 50 MPa. The set temperature is slightly higher than $T_c$, which ensures the existence of an energy barrier to describe the nature of first-order MT yet allows a small strain fluctuation to jump over. The sequential process of nucleation and growth of martensite is shown in Fig. 1, where martensite is depicted by the distribution of the deviatoric strain $e_2$ at different times during the isothermal simulation. The displacement fluctuations give rise to local inhomogeneities of the deviatoric strain at the initial stage of the isothermal simulation that trigger the onset of a fine-scale assembly of cross-hatched tweeds (Fig. 1-b). Differing from single crystals [62], the tweeds exhibit different stretch directions in individual grains. Note that tweeds similar to those in Fig. 1-b can also develop above $T_m$ in the absence of applied stresses. The local distortion of deviatoric strain caused by the local stress field is hence accumulated at the grain boundaries with larger misorientation, as those between blue and red grains in Fig.1-b, indicating that the free energy minima have been momentarily shifted toward the grain boundaries due to the synergistic contribution of the grain boundary energy and the strain gradients.

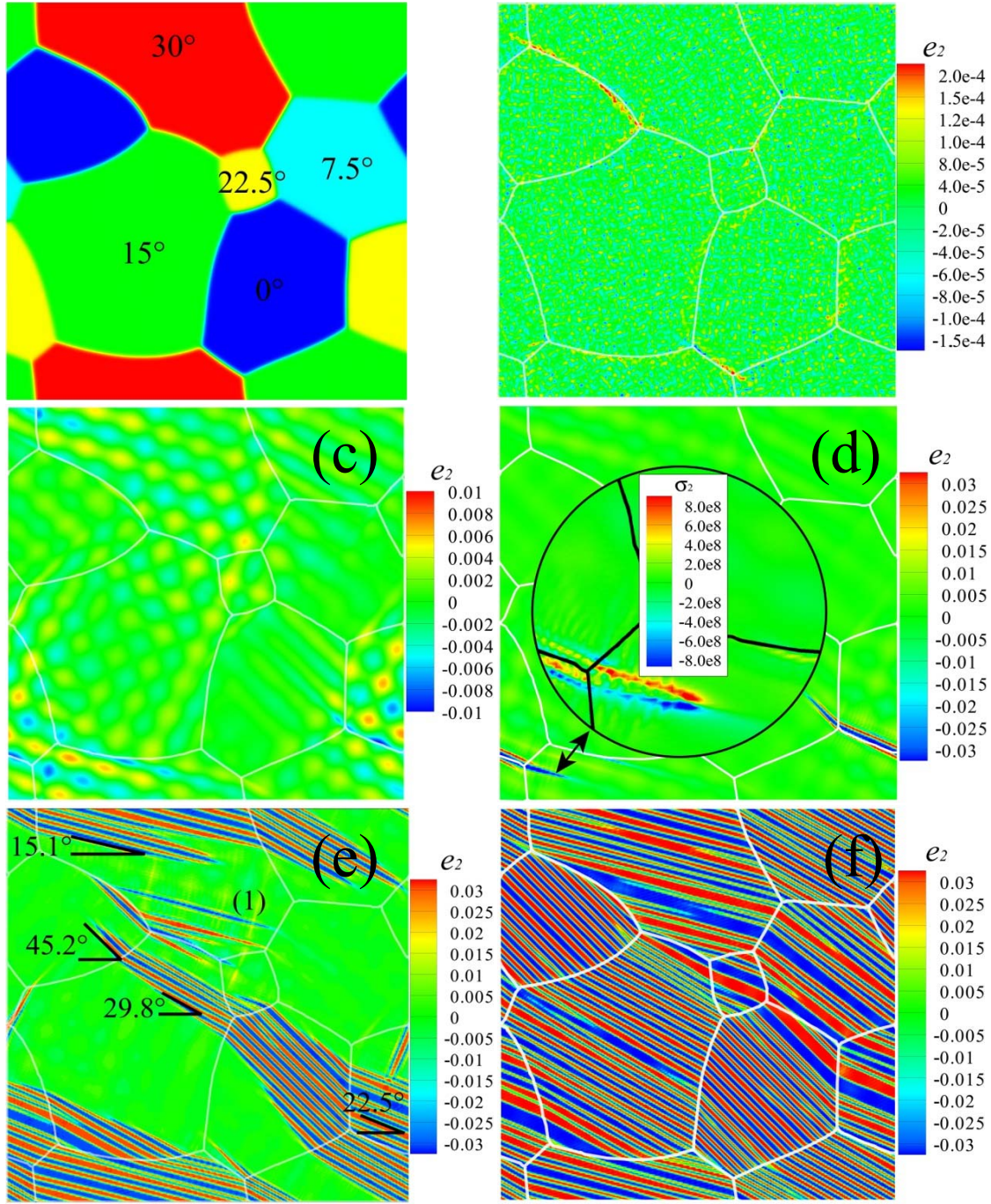

Figure 1. (a) Spatial distribution of the grain orientation angle as indicated in each grain (b)-(f) Martensitic nucleation and growth in polycrystals, as shown by the deviatoric strain $e_2$ at 275K under an applied stress $\sigma_{app}$= 50MPa, at time steps of 800, 12000, 92000, 94800 and 104000, respectively. The red and blue domains represent two martensite variants while the green domain is austenite. The grain boundaries are displayed in white.

For $\sigma_{app}$ = 50 MPa, no concentration of deviatoric strain is found in the vicinity of triple junctions because the strain gradients stemming from individual grain boundaries cancel each other and lead to a relatively more homogenous strain field at the triple junction. The fine tweeds inside the grains and deviatoric strain distortion at grain boundaries are fading away with time, and the system responds to the local stress field with a metastable microstructure exhibiting a long-range straining pattern (Fig. 1-c), denominated 'postcursor' [63]. The postcursors, modulated partially transformed martensites, show the same texturing pattern as what the subsequent twins would have. They arise from a balance between the nonlinearity of transformation energy represented by the anharmonic terms in Eq. (4) and nonlocality stemming from the contribution of strain gradients. This modulated metastable equilibrium will be broken when the reduction of local energy due to the stress released in martensite domains cannot be compensated by the increase of energy originating from the steeper strain gradient with respect to the spatial coordinates, and the total energy is progressively reduced to the stable energetic wells of martensite [63]. Therefore, the postcursors can play the role of potential embryo of martensite inside the grains, as shown by the martensite plate marked (1) in Fig.1-e, which grows from an intragranular postcursor. Simultaneously, the strain distortion at the grain boundary can assist nucleation. If the grain boundary is parallel to one of the postcursors, the first stable martensite plate will form preferentially along the grain boundary, as shown in Fig.1-d. As soon as the first stable martensite plate forms (see the blue variant and the local deviatoric stress $\sigma_2(\vec{r}) = (\sigma_{xx}(\vec{r}) - \sigma_{yy}(\vec{r}))/\sqrt{2}$ in inset of Fig 1-d), it grows rapidly in length following the tracks of postcursors due to the large local stress and the curvature at the tip. When the growing front of the martensite plates crosses the triple junction, the martensite plate penetrates into the grain and is aligned with the favored orientation. However, the coarsening of the blue variant in the direction perpendicular to the habit plane is momentarily arrested by the modified local energy around the plate, and gives rise to a cascade of pairing red variants, i.e. autocatalysis phenomenon. It should be noted that branched martensite plates are hardly found due to the limitations of the 2D model that allows only two different twin-related variants. The branched martensite plates are only found when the grain boundary is located between the orientations of the postcursors in the neighboring grains (see inset in Fig. 3-b). This is different to the richer accommodated patterns consisting of all 24 Kurdjumov–Sachs orientation variants widely observed in the experiments for other Fe-C based alloys [64]. Figure 1-f illustrates the morphology of full martensite after long-time simulation. The alternative alignments of twinned red and blue martensites are in agreement with the kink solution of the solitary wave in Ref. [42]. The width and energy density of the twin boundary shown in Fig.1-f correspond to about 2-3 nm and 0.45J/m$^2$, respectively. Although the simulated twin boundary energy density agrees well with that of stationary analysis, the width of the twin boundary in the simulation is three times broader than the several atomic diameters observed from TEM image. It is an inherent consequence of the diffuse interfaces utilized in Landau modeling. In this work, the gradient coefficient was selected by considering a compromise between the twin boundary width and the size of the grains.

### *3.2 Heterogeneous nucleation under an applied stress of 500 MPa*

The evolution of microstructure under $\sigma_{app}$ = 500 MPa shows similar features (Fig. 2-a) at the early stages of the simulation as those for 50 MPa: the tweed pattern develops within the grains and the strain distortions appear at grain boundaries, but both occur more rapidly because of the higher stresses. The accumulation of the deviatoric strain raised by the local stress is supposed to smear out or weaken the effect of postcursors. At the same time, the interplaying of elastic transformation energy and lattice curvature drives the 'twinned' strain distortion to split into 'diploes' (indicated by the arrows in Fig.2-a) and to be concentrated at the triple junctions and high angle grain boundaries, both of which serve as the preferential sites for martensitic nucleation and growth, as shown in Fig.2-b. Three martensite plates (marked (1), (2) and (3) in Fig. 2-c) are nucleated at the triple junctions, and grow into the twins aligned with their optimum martensite orientations in individual grains. For instance, the pairs of needle-like martensites in the grains with the orientation θ = 22.5° (i.e. the grains in yellow in Fig. 1-a) align along the 67.5° with respect to the horizontal axis of global reference.

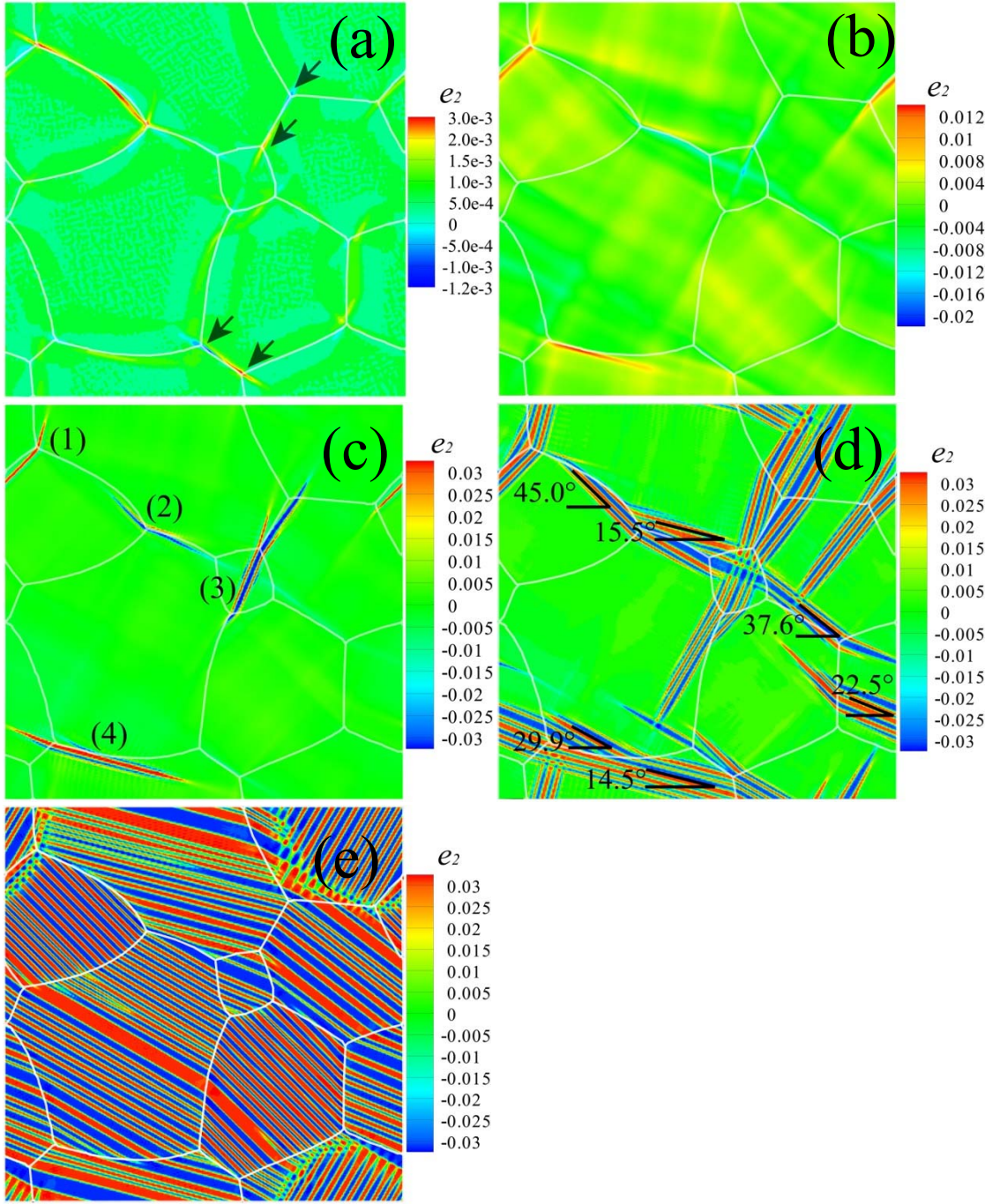

Figure 2. Martensitic nucleation and growth in polycrystals, as shown by the deviatoric strain $e_2$ at 275K under an applied stress $\sigma_{app}$=500MPa at time steps of 800 (a), 5200 (b), 6400 (c), 8000 (d), and 28000 (e), respectively. The red and blue domains represent two martensite variants while the green domain is austenite. The grain boundaries are displayed in white.

A different scenario is found in the twinned martensite plates marked (4) in Fig. 2-c. They nucleate and grow assisted by the high angle grain boundaries, and the leading plate in red propagates beyond the neighbor triple junction without apparently changing orientation. However, it accommodates itself later to match the surrounding grain misorientation by subtly rotating the residing twin boundary to form the full martensitic morphology (Fig.2-e).

*3.3 The effect of applied stress on heterogeneous nucleation at grain boundaries*

In Fig.3, the deviatoric strain as a function of the distance to a high angle grain boundary is plotted for different applied stresses and times at $T$ = 275 K. Although the high angle grain boundaries can be intuitively considered analogous to 2D planar defects in Refs [41-42], we must clarify that our modeling has its roots in the dynamical nucleation developed by Clapp [65], Cao [40-41], and Gooding [42], *et al.* The hyperbolic-type profiles of the deviatoric strain are consistent with the analytical results of Cao [41] who introduced a predefined stress field as a martensite embryo for twin. The deviatoric strains raised by the small local stresses rapidly accumulate within the first hundreds of steps (see inset of Fig. 3 with $\sigma_{app}$= 20-100 MPa). However, they are one to two orders of magnitude smaller than the threshold of 'spinodal strain' to serve as steady martensite embryos to form stable martensite variants. The strain profile at $\sigma_{app}$= 500 MPa and at time step of 6400 (in the direction perpendicular to the twin plane 4# in Fig. 2-c) exhibits the oscillation between the positive/negative deviatoric strains of two rectangular martensite variants, demonstrating that the local stress has driven the martensitic nuclei into stable plates. Beyond the analytical solution, our results prove that the spatiotemporal evolution is greatly dependent on the strain gradient. In contrast to the predefined stress solution [41], the asymmetry of the profile with $\sigma_{app}$ = 500 MPa and at time step of 2800 indicates that the strain distortion is influenced by a curvature driving modulation (through either strain gradients or displacement gradients) and then is concentrated at the triple junctions. As a result, more martensites nucleate at the triple junctions under $\sigma_{app}$= 500 MPa (see martensites 1#-3# in Fig. 2-c) than at high angle grain boundaries. All the above findings reveal that intragranular nucleation around grain boundaries prevails over other nucleation modes under small applied stress, whereas nucleation at triple junctions and high angle grain boundaries is the dominant mode under moderate applied stress. These results support the argument of Clapp [39], who states that high angle grain boundaries are the favored nucleation sites if the free surface and incoherent grain boundaries are absent. It should be further emphasized that the triple junctions are more suitable nucleation sites than high angle grain boundaries in the case of the relative large local stresses.

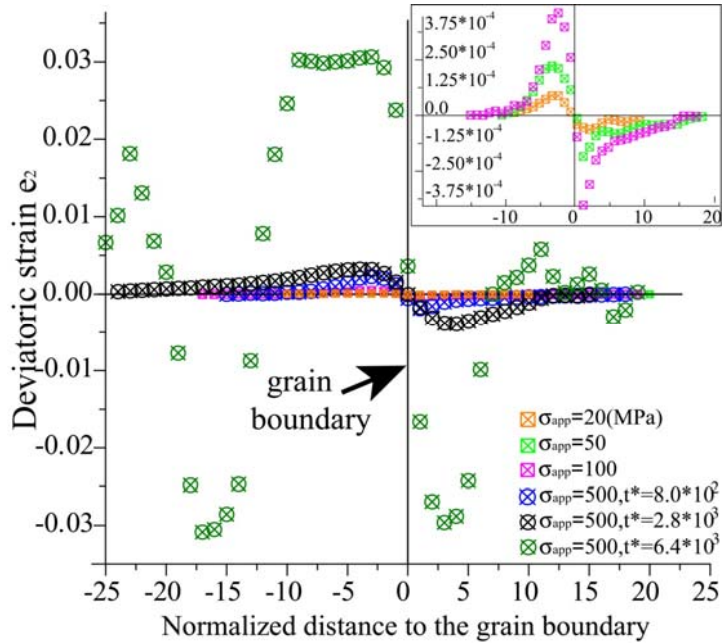

Figure 3. Evolution of the deviatoric strain $e_2$ as a function of the distance to a high angle grain boundaries at the time step of 800 under different applied stresses $\sigma_{app}$=20, 50, 100 and 500 MPa, respectively, and at later times of 2800 and 6400 steps with $\sigma_{app}$= 500 MPa .

## 3.4 The effect of quenching temperatures on heterogeneous nucleation at grain boundaries

The effect of quenching temperatures on the Landau energy landscape and the transformation pathways is addressed in Fig. 4, which illustrates the morphology of the principal martensitic plates at different temperatures under $\sigma_{app}$=100 MPa. When quenching the system to a low temperature $T = 250K <T_c$, the martensite plates prefer to form and grow at the triple junctions prior to the boost of homogenous decomposition inside grains. Quenching to $T_c<T$=275 K$<T_m$, in contrast, leads to a diversity of martensitic nucleation sites depending on the localized geometric, energetic and dynamic conditions, when the stress fields localized at the postcursors, grain boundary and triple junctions are comparable in magnitude. The groups of stripe-like martensites (marked (1) in Fig. 4-b) nucleate to grow intragranularly following the tracks of postcursors. On the contrary, the red variant nucleates at the high angle grain boundary, proceeds along the prestrained postcursors after crossing

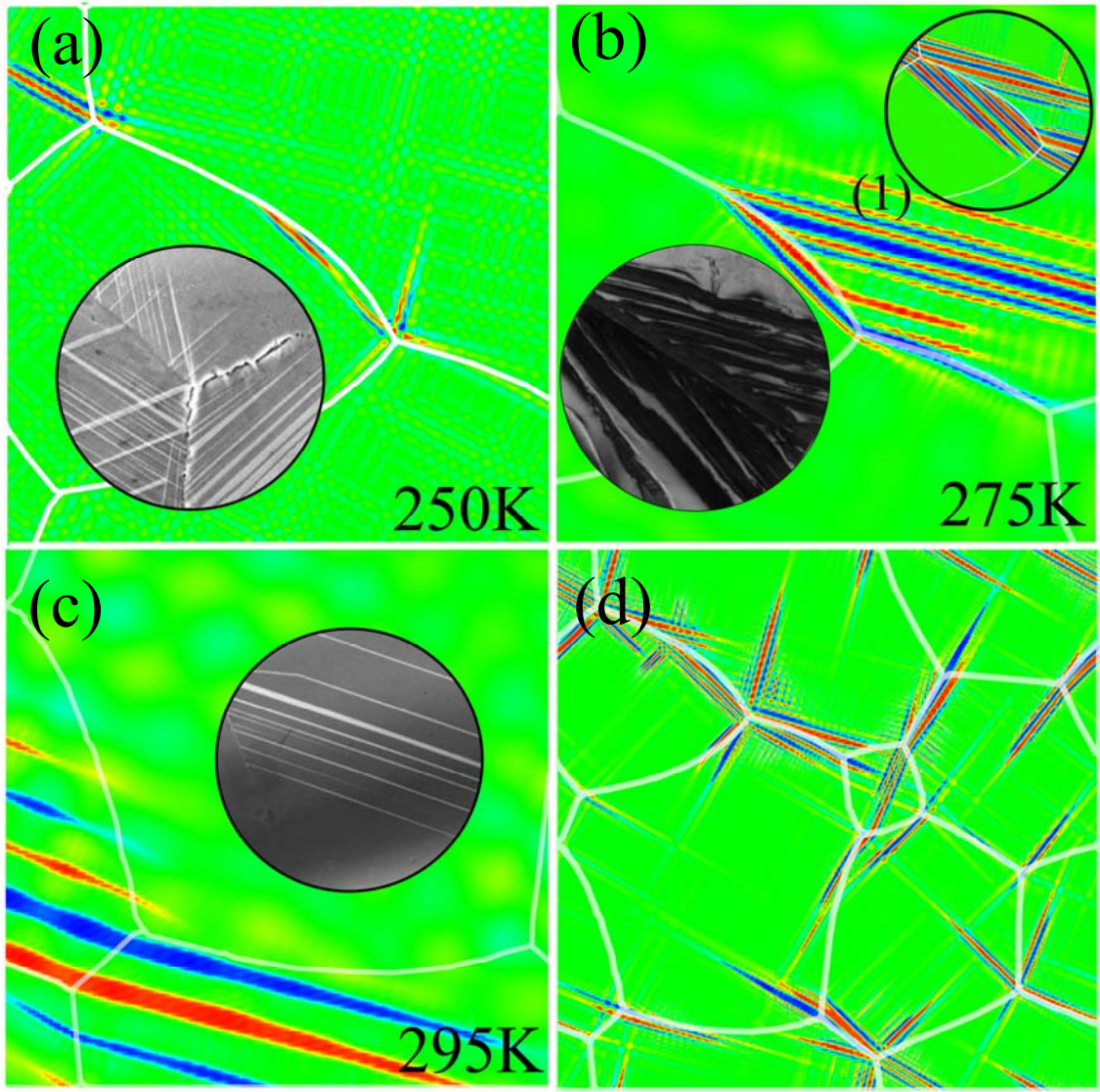

Figure 4. Morphology of the initial stabilized martensitic plates in polycrystals undergoing isothermal transformation after quenching to 250K(a), 275K(b) and 295K(c) and continuous cooling at the rate of 0.01K/step from 300K to 100K(d), as shown by the deviatoric strain $e_2$ under an applied stress $\sigma_{app}$= 100 MPa

the grain boundary, but leaves with a kink; while the blue martensite plate nucleated at the triple junction and accommodates itself to match the optimal orientations in the neighboring grains. The later morphology at the very high angle grain boundary (see the top-right inset in Fig. 4-b) is consistent with the coarse lenticular martensites observed in Fe-based materials [64]. On quenching to $T = T_m$ =295K, the morphology exhibits a heterogeneous structure containing austenite-martensite alternated stripes in the grains (Fig. 4-c). This process takes place via the nucleation and growth relying on the preexisting postcursors with a morphology much like the stress-induced martensite at room temperature ($>T_m$) for certain cubic-to-monoclinic II shape memory alloys [66]. Note that since the symmetry breaking of cubic-to-monoclinic II MT contains the elements of cubic-to-tetragonal, a specific 2D projection of the microstructure of the monoclinic II martensites in a fine wire sample could sustain the lamella features in tetragonal martensites [67]. The formation of a single martensite variant in equilibrium with the austenite is energetically allowed at $T_m$, and even at higher temperature, if the local stress is above a threshold. Nevertheless, it is constrained by geometrical compatibility in 3D bulk materials. So far, our simulations have demonstrated that the martensitic nucleation modes (sites) are sensitive to the quenching temperature. The primary nucleation sites vary from triple junctions to high angle grain boundaries to postcursors as the quenching temperature increases. Thus, it can be inferred that a continuous cooling to a lower temperature is perfectly adequate to yield the formation of stable martensite plates at the triple junctions, as proven in Fig. 4d.

*3.5 Phase diagram of nucleation modes*

Most of the discoveries by our simulations can be summarized in the phase diagram of the Fe-Pd ferroelastics presented in Fig. 5. The phase diagram looks similar to a conventional TTT diagram, and conveys the information about the principally favored mode of martensitic nucleation under particular combination of quenching temperature and applied stress. It is still noteworthy that the nucleation and growth of martensite is actually governed by the local stresses rather than by the applied stress. However, the applied stresses can be easily used to control processing in an industrial environment. The phase diagram manifests that nucleation at the postcursors dominates at high quenching temperatures. As the quenching temperature decreases or the applied stress increases, nucleation at the postcursors is replaced by four distinct mixed modes, namely, grain boundary + postcursor, triple junction + postcursor, hybrid (i.e. all sorts of nucleation sites are possible) and triple junction + grain boundary. Only the postcursors can lead by themselves to the development of a nucleation site, while the grain boundaries and triple junctions have to cooperate to become viable nucleation sites, see Fig.5. It is worth pointing out that our results have been obtained assuming that the material was elastic and continuous. Thus, other aspects associated with grain boundary nucleation such as plasticity, incoherent interfaces and defect cores have not been considered. The incorporation of these issues into the simulation will be necessary to include more physics in the model and enhance the predictive capabilities of this strategy.

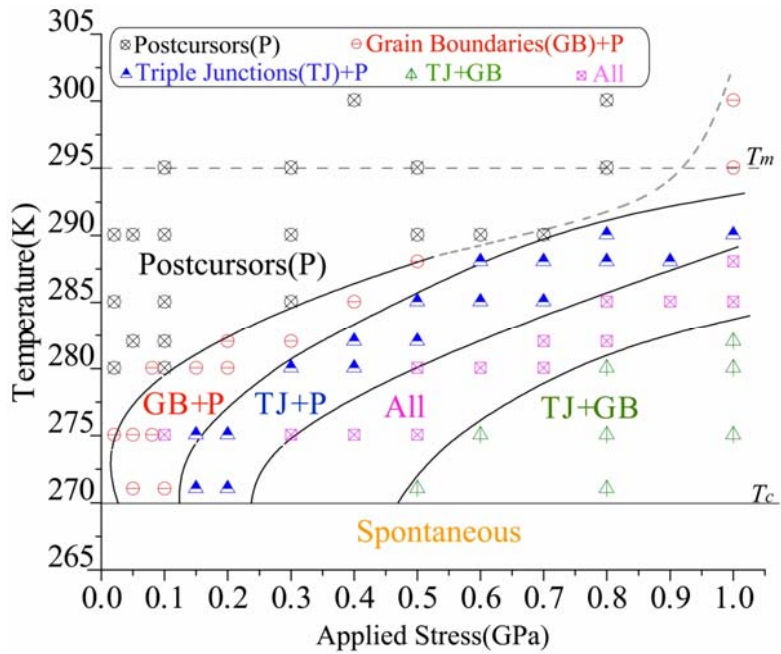

Figure 5. Phase diagram of the nucleation modes of martensite in the Fe-Pd ferroelastic polycrystals.

## 4. Conclusion

In summary, we investigate the dynamical nucleation of martensite in polycrystals by means of LR dynamics and Landau energetics. Avoiding the detail of the defect cores, the simulations capture the heterogeneous processes of dynamical martensitic nucleation under the local stress field, and are able to predict the phase diagram of martensitic nucleation. With the example of Fe-Pd ferroelastic polycrystals, the phase diagram illustrates that postcursors, high angle grain boundaries and triple junctions serve as the preferential heterogeneous nucleation sites depending on the loading and cooling conditions. The information, presented in the form of a phase diagram like Fig. 5, specify how the quenching temperature and applied stress conditions can be combined to activate particular mechanisms of martensitic nucleation and growth, leading to complex microstructures in order to meet microstructural design goals.

*Acknowledgement*

This investigation was supported by the European Research Council (ERC) under the European Union's Horizon 2020 research and innovation programme (Advanced Grant VIRMETAL, grant agreement No. 669141), and also supported by European Union's Research Fund for Coal and Steel (RFCS) under the Grant Agreement RFSR-CT-2011-00017. GX wishes to acknowledge the financial support by the Chinese Scholarship Council (CSC). We are grateful to Prof. Arnaud Weck for discussion.